\documentstyle[12pt,preprint,aps,tighten,epsfig,floats]{revtex}

\def\beq{\begin{equation}}
\def\eeq{\end{equation}}
\def\beqa{\begin{eqnarray}}
\def\eeqa{\end{eqnarray}}
\def\be{\begin{equation}}
\def\ee{\end{equation}}
\def\bea{\begin{eqnarray}}
\def\eea{\end{eqnarray}}
\begin{document}

\title{Dimension-eight Operators in the Weak OPE}
\author{Vincenzo Cirigliano, John F. Donoghue and Eugene Golowich} 
\address{Department of Physics and Astronomy, University of Massachusetts\\
Amherst, MA 01003 USA \\ 
vincenzo@het2.physics.umass.edu
donoghue@physics.umass.edu \\
golowich@physics.umass.edu \\}
\maketitle
\thispagestyle{empty}
\setcounter{page}{0}
\begin{abstract}
\noindent We argue that there is a potential flaw in the standard treatment of
weak decay amplitudes, including that of $\epsilon' / \epsilon$.  We
show that (contrary to conventional wisdom) dimension-eight operators
{\em do} contribute to weak amplitudes, at order $G_{F} \alpha_s$ 
and without $1/M_W^2$ suppression.  We demonstrate the existence 
of these operators through the use of a simple weak hamiltonian. 
Their contribution appears in different places depending on which scheme
is adopted in performing the OPE.  If one performs a complete separation
of short and long distance physics within a cutoff scheme,
dimension-eight operators occur in the weak hamiltonian at order
$G_{F} \alpha_s / \mu^2$, $\mu$ being the separating scale. 
However, in an $\overline{\rm MS}$ renormalization scheme for the OPE 
the dimension-eight operators do not appear explicitly in the 
hamiltonian at order $G_{F} \alpha_s$. 
In this case, matrix elements must include physics above the
scale $\mu$, and it is here that dimension eight effects enter.
The use of a cutoff scheme (especially
quark model methods) for the calculation of the matrix elements of
dimension-six operators is inconsistent with  
$\overline{\rm MS}$ unless there is careful matching including 
dimension-eight operators.  The
contribution of dimension-eight operators can be minimized by working
at large enough values of the scale $\mu$.  We find from 
sum rule methods that the contribution of dimension-eight operators 
to the dimension-six operator ${\cal Q}_7^{(6)}$ is at the 100$\%$ level for 
$\mu = 1.5$ GeV.  This suggests that presently available values of 
$\mu$ are too low to justify the neglect of these effects.  Finally, 
we display the dimension-eight operators which appear within the 
Standard Model at one loop. 
\end{abstract}

\vspace{1.0in}

\section{Introduction}

The starting point for the study of nonleptonic weak transitions is the
analysis of short-distance effects using perturbative QCD. The results are
expressed using the Operator Product Expansion (OPE) as a series
of local operators. In practice, only operators of dimension six 
are considered. We will argue that operators of dimension eight are 
also relevant, and that most previous analyses of nonleptonic 
amplitudes must be reconsidered.

Nonleptonic weak amplitudes represent probably the most difficult 
calculations in QCD. Since the W-boson propagator is a constant 
up to $Q \sim M_W$ in momentum space , the amplitude is sensitive to 
strong interaction physics at {\em all} energy scales.  Therefore, 
one must control simultaneously the very low, intermediate and 
high energy portions of the calculation. There are two key ideas, 
both introduced by Wilson\cite{Wilson1,Wilson2,Wilson3,Wilson4}, 
that are used in this regard. One consists of
separating the different energy scales and integrating out those effects
from high energy. This yields an effective low energy theory with modified
interactions. The second is the tool for doing this - the
Operator Product Expansion (OPE) - in 
which the effects of high energy are replaced by 
local operators ordered according to increasing dimension.
The latter can be applied at any scale, and as we reduce this scale
we successively integrate out more and more physics, thereby changing
the coefficients of the operators.  Specifically, one can consider 
the physics above and below some energy scale $\mu$. Throughout 
the paper, we will refer to $\mu$ as the {\it separation scale}.  
For example, if one takes the complete set of all local 
dimension-$d$ operators\footnote{Throughout this 
paper, we explicitly display the operator dimension as a 
superscript, {\it e.g.} ${\cal O}^{(d)}$.} 
 $\{ {\cal Q}_i^{(d)} \}$ with the right 
quantum numbers, the operator 
product expansion tells us that a $\Delta S = 1$ amplitude can 
be written to leading order in dimension as
\beq
\langle {\cal H}_W^{(\Delta S = 1)} \rangle  \ = \ {G_F \over \sqrt{2}} ~
V_{us}V_{ud}^* \sum_d \sum_{i} ~{\cal C}_i^{(d)} (\mu)~ \langle 
{\cal Q}_i^{(d)} \rangle_\mu \ \ .
\label{dim1}
\eeq 
Here the $\{ {\cal C}_i^{(d)} (\mu) \}$ are coefficients which describe the 
short distance physics with $Q \ge \mu$. The subscript 
`$\mu$' on the operator matrix element indicates that the matrix 
element is to include all physics up to the energy scale 
$\mu$ ({\it i.e.} with $Q \le \mu$).  The short distance 
OPE does not by itself solve the problem of nonleptonic 
amplitudes. However, it does tell us that the remaining task 
is to calculate the low energy matrix elements of local operators.

\begin{figure}
\vskip .1cm
\hskip 3.0cm
\psfig{figure=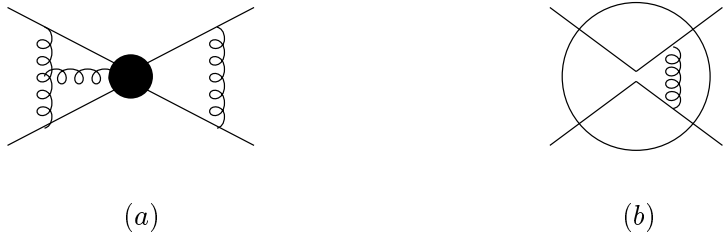,height=1.2in}
\caption{Scales in the weak transitions: (a) Long range, 
(b) Short range.\hfill 
\label{fig:dim8fig1}}
\end{figure}

An example of this appears in Fig.~\ref{fig:dim8fig1}, where 
the long-distance (low-energy) and short-distance (high-energy) 
parts of a nonleptonic weak transition are depicted separately.  
Let the separation scale be $\mu$. 
Then the gluons shown in Fig.~\ref{fig:dim8fig1}(a) have $Q \le \mu$ 
and are associated with long-distance propagation.  The blackened 
disc denotes all the (short-distance) effects with $Q \ge \mu$.  
We next look into the short-distance regime via 
Fig.~\ref{fig:dim8fig1}(b).  Now there are only hard gluons with 
$Q \ge \mu$ which propagate over short-distances. 
If the separation scale $\mu$ is large enough, the physics at 
the higher energies can be analyzed with perturbative QCD.  Because 
of this, the short-distance effects will appear local when 
viewed by low energy probes.  Also, observe in 
Fig.~\ref{fig:dim8fig1}(b) that the W-boson mass has been 
taken to infinity and so the process shown there corresponds to 
the range $\mu \le Q \ll M_W$.  

The most obvious candidates for the basis of weak nonleptonic 
operators are those with dimension six, formed as the product 
of two currents, 
\beq
{\bar q}_1 \Gamma q_2 ~ {\bar q}_3 \Gamma q_4 \ \ .
\label{dim2}
\eeq 
Since the hamiltonian is the product of two weak currents, the 
coefficients of these operators are dimensionless. Next will come 
the operators of dimension eight, some examples of which are 
\beq
{\bar q}_1 \Gamma {\cal D}_\mu q_2~ {\bar q}_3 
\Gamma {\cal D}^\mu q_4 \ \ ,
\label{dim3}
\eeq
where ${\cal D}_\mu q_i$ is the covariant derivative and 
\beq
f_{abc}{\bar q}_1 {\lambda^a \over 2} \Gamma^\mu q_2~
{\bar q}_3 {\lambda^b \over 2} 
\Gamma^\nu q_4 F^c_{\mu\nu} \ \ , 
\label{dim4}
\eeq
where $F^c_{\mu\nu}$ is the gluon field-strength tensor.  The 
coefficients of these operators have engineering dimension 
(Energy)$^{-2}$. In the original papers\cite{Gaillard,Altarelli} 
on the weak interaction 
OPE it was stated explicitly that the coefficients of dimension-eight 
operators are of order $1/M_W^2$, and this has been accepted 
ever since. All the current treatments consider only operators of 
dimension six\cite{Ciuchini,Buras,overviews}.  
We will show that the correct procedure is more subtle.  

Specifically, we will first consider 
the situation where one has a true separation of 
scales in the fashion outlined above. In this case:
\begin{enumerate}
\item  Dimension-eight operators enter 
the weak OPE at order $1/\mu^2$ and not $1/M_W^2$.
\item In the one calculable example that we know about, such effects 
continue to be significant even above the scale $\mu\sim 2$~GeV. 
Since most present calculations are performed with $\mu \sim 0.7 \to 2$~GeV,
dimension-eight effects will likely affect the results of past work.
\item Most generally, the dimension-eight effect can appear in both the 
coefficient functions and the matrix elements.  The 
relative amount of each depends on how one implements 
the division of physics at the scale $\mu$ and amounts to 
a `separation scheme' dependence. 
\end{enumerate}
However, in the process of demonstrating these 
points, we will also see that dimensional regularization
does not accomplish the separation of physics above and below the
scale $\mu$. Since dimensional regularization is by far the 
easiest calculational scheme, we study the structure of the OPE in
such a scheme. We find:
\begin{enumerate}
\item[4.] Matrix 
element evaluations in dimensional regularization must be 
sensitive to energies {\em above} the scale $\mu$.
\item[5.] In this case, the effects of dimension-eight operators 
appear fully within
the matrix elements of dimension-six operators.
\item[6.] Mixed evaluations, in which one calculates the coefficients 
using dimensional regularization and the matrix elements using a 
form of a cutoff, are inherently inconsistent. 
Most past calculations fall in this category.
\item[7.] The influence of dimension-eight effects can be controlled 
by working at sufficiently large $\mu$.   Further work will be 
required to understand just how large $\mu$ must be to 
achieve a given precision.
\end{enumerate}

This work has two basic parts. In the first, we use an explicit 
analytic calculation to illustrate the properties of 
dimension-eight operator effects in a weak amplitude. This will
provide a demonstration of the above points. 
In the second part, we calculate the relevant dimension-eight 
operators for the Standard Model $\Delta S = 1$ weak hamiltonian 
in a particular separation scheme. This will allow the exploration 
of the size of such effects, provided the operator
matrix elements can be evaluated on the lattice.

\section{An explicit example}

Rather than deal with the usual weak hamiltonian, we start with a similar 
but distinct operator that has simplified properties and allows us
to demonstrate analytically the existence and properties of the 
dimension-eight operators. This hamiltonian\cite{dg,others} 
contains one 
left-handed and one right-handed current instead of the usual Standard 
Model hamiltonian in which both currents are left-handed. Specifically
we define\footnote{We omit CKM dependence 
in the operator ${\cal H}_{\rm LR}$ and define the chiral matrices 
$\Gamma^\mu_{L \atop R} \equiv \gamma^\mu (1 \pm \gamma_5)$.} 
 \beqa
{\cal H}_{\rm LR}  &\equiv& 
{g_2^2 \over 8} \int d^4x \ 
{\cal D}_{\mu\nu}(x,M_W^2) ~J^{\mu\nu}(x) \ \ ,
\nonumber \\
J^{\mu\nu}(x) &\equiv& {1\over 2} 
T\left[ {\bar d}(x) \Gamma^\mu_L u (x) ~
{\bar u}(0) \Gamma^\nu_R s (0) \right]
\nonumber \\
&=& {1\over 2} 
T\left[ \big(V^\mu_{1 - i2} (x) + A^\mu_{1 - i2} (x)\big) ~
\big(V^\nu_{4 + i5} (0) - A^\nu_{4 + i5} (0) \big)
\right] \ \ ,
\label{va}
\eeqa
where ${\cal D}_{\mu\nu}$ is the $W$-boson propagator and $V^\mu_a$, 
$A^\mu_a \ (a = 1,\dots 8)$ are the flavor-octet vector, 
axialvector currents.  

The reason why this hamiltonian provides a useful example is that in the  
chiral limit its matrix elements are related to vacuum matrix elements, and 
we may therefore take advantage of what is known about the associated 
vacuum polarization functions.~\cite{svz,sumrule}  For example,
the K-to-pi matrix element 
\beq
{\cal M}(p) =  \langle \pi^- (p)| 
{\cal H}_{\rm LR} | K^{-} (p) \rangle
\label{dim5}
\eeq
is given in the chiral limit of zero momentum and vanishing 
light-quark masses by the vacuum matrix element
\beq
{\cal M} \equiv \lim_{p=0}{\cal M}(p) 
= {g_2^2 \over 16 F_\pi^2} \int d^4x \ {\cal D}(x,M_W^2) ~
\langle 0 |T\left( V^\mu_3  (x) V_{\mu,3}  (0) - 
A^\mu_3 (x) A_{\mu, 3} (0)\right) | 0 \rangle \ \ .
\label{va3}
\eeq
Two of us have recently studied the amplitude ${\cal M}$ 
and more details on its properties can be found in Ref.~\cite{dg}. 
Here we will display those features useful for 
understanding the role of dimension-eight operators.

One can perform an operator product expansion on the hamiltonian
${\cal H}_{\rm LR}$ in the usual fashion.  Including only the
dimension-six operators one finds\footnote{We stress that $\{
{\cal O}^{(6)}_k \}$ and $\{ c^{(6)}_k \}$ of this section 
are distinct from $\{
{\cal Q}^{(6)}_i \}$ and $\{ {\cal C}^{(6)}_i \}$ of
Eq.~(\ref{dim1}).}  \beq {\cal M} \simeq {G_F \over 2 \sqrt{2}F_\pi^2}
\left[ c_1^{(6)} (\mu) \langle {\cal O}^{(6)}_1 \rangle_{\mu} + c_8^{(6)}
(\mu) \langle {\cal O}^{(6)}_8 \rangle_{\mu} \right] \ \ .
\label{sd7}
\eeq
The operator basis consists of two left-right operators 
${\cal O}^{(6)}_1$, ${\cal O}^{(6)}_8$ which have respectively color-singlet 
and color-octet structure, 
\beqa
{\cal O}^{(6)}_1 &\equiv& {\bar q} \gamma_\mu {\tau_3 \over 2} q
~{\bar q} \gamma^\mu {\tau_3 \over 2} q - 
{\bar q} \gamma_\mu \gamma_5 {\tau_3 \over 2} q 
~{\bar q} \gamma^\mu \gamma_5 {\tau_3 \over 2} q \ \ ,
\nonumber \\
{\cal O}^{(6)}_8 &\equiv& {\bar q} \gamma_\mu \lambda^a
{\tau_3 \over 2} q
~{\bar q} \gamma^\mu \lambda^a {\tau_3 \over 2} q - 
{\bar q} \gamma_\mu \gamma_5 \lambda^a {\tau_3 \over 2} q
~{\bar q} \gamma^\mu \gamma_5 \lambda^a {\tau_3 \over 2} q
\ \ .
\label{sd4}
\eeqa
In the above, $q = u,d,s$, $\tau_3$ is a Pauli (flavor) matrix, 
$\{ \lambda^a \}$ are the Gell~Mann color 
matrices and the subscripts on ${\cal O}^{(6)}_1$, ${\cal O}^{(6)}_8$ 
refer to the color carried by their currents.
The coefficient functions, including renormalization group summation, are
\beqa
c_1^{(6)} (\mu) &=& {1 \over 9} \left[ \left( {\alpha_s (\mu) 
\over \alpha_s (M_W)} \right)^{8/9} + 8 
\left( {\alpha_s (\mu) \over \alpha_s (M_W)} 
\right)^{-1/9} \right] \ \ , 
\nonumber \\
c_8^{(6)} (\mu) &=& {1 \over 6} \left[ \left( {\alpha_s (\mu) 
\over \alpha_s (M_W)} \right)^{8/9} - \left( 
{\alpha_s (\mu) \over \alpha_s (M_W)} \right)^{-1/9} \right] \ \ ,
\label{sd8}
\eeqa
with 
\beq
\alpha_s (\mu) = \left[ 1 + 9 {\alpha_s (\mu) \over 4 \pi}
\ln \left( {M_W^2 \over \mu^2} \right) \right] \alpha_s (M_W) \ \ .
\label{sd9}
\eeq
For our purpose it is sufficient to work with 
an expansion of Eq.~(\ref{sd7}) through first order 
in $\alpha_s (\mu)$, 
\beq
{\cal M} \simeq {G_F \over 2 \sqrt{2}F_\pi^2} 
\bigg[ \langle {\cal O}^{(6)}_1 \rangle_{\mu} \ 
+ \ { 3 \over 8 \pi} \ln \left( {M_W^2 \over \mu^2} 
\right) \langle \alpha_s {\cal O}^{(6)}_8 \rangle_{\mu} \ + \ldots \bigg] \ \ .
\label{sd10}
\eeq
Our goal is next to carry out an explicit evaluation
of ${\cal M}$ and to demonstrate that dimension-eight operators 
appear in addition to those of dimension six.

\subsection{The Presence of Dimension-eight Operators}

Let us analyze the vacuum matrix element that appears 
in Eq.~(\ref{va3}) in terms of the 
vacuum polarization function
\beqa
& & i \int d^4 x\ e^{i q \cdot x} 
\langle 0 |T\left( V^\mu_3  (x) V^\nu_3 (0) - 
A^\mu_3 (x) A^\nu_3 (0)\right) | 0 \rangle 
\nonumber \\
& & \phantom{xxxxx} = (q^\mu q^\nu - q^2 g^{\mu\nu} ) 
( \Pi_{V,3} - \Pi_{A,3} )(q^2) - q^\mu q^\nu \Pi_{A,3}^{(0)}(q^2) 
\ \ .
\label{r0}
\eeqa
Using this we transform the spatial integral in Eq.~(\ref{va3})
to momentum space, 
\beq
{\cal M} =  {3 G_F M_W^2 \over 32 \sqrt{2}\pi^2 
F_\pi^2} \int_0^{\infty} dQ^2 \ {Q^4 \over Q^2 + M_W^2} 
\left[ \Pi_{V,3} (Q^2) - \Pi_{A,3} (Q^2) \right] \ \ .
\label{full}
\eeq
Next we implement the separation of scales for the operator 
product expansion. We do this by applying a cutoff at $Q^2=\mu^2$ and 
using the OPE in the high-energy/short-distance portion.

Thus consider a partition of ${\cal M}$ 
characterized by the scale $\mu$, 
\beq
{\cal M} = {\cal M}_<(\mu) + {\cal M}_>(\mu) \ \ ,
\label{c6}
\eeq
where ${\cal M}_<(\mu)$ and ${\cal M}_>(\mu)$ are 
dependent respectively on contributions with $Q < \mu$ and 
$Q > \mu$.  We obtain then for the low energy portion, 
\beqa
{\cal M}_< (\mu) &=&  {3 G_F M_W^2 \over 32 \sqrt{2}\pi^2 F_\pi^2} 
\int_0^{\mu^2} dQ^2 \ {Q^4 \over Q^2 + M_W^2} 
\left[ \Pi_{V,3} (Q^2) - \Pi_{A,3} (Q^2) \right] 
\nonumber \\
&=& {3 G_F \over 32 \sqrt{2}\pi^2 
F_\pi^2} \int_0^{\mu^2} dQ^2 \ Q^4 
\left[ \Pi_{V,3} (Q^2) - \Pi_{A,3} (Q^2) \right] 
+ {\cal O}(\mu^2 / M_W^2) \ \ .
\label{va6} 
\eeqa
This cut-off is well-defined as it refers to the external 
momentum of a gauge-invariant amplitude.
We have shown in Ref.~\cite{dg} 
that this relation serves as a definition 
of the vacuum matrix element for the local operator 
${\cal O}^{(6)}_1$ at the scale $\mu$ with a momentum-cutoff scheme
(denoted as `(c.o.)'),
\beq
\langle {\cal O}^{(6)}_1 \rangle_\mu^{\rm (c.o.)} = {3 \over 16 \pi^2} 
\int_0^{\mu^2} dQ^2 ~Q^4 \left[ \Pi_{V,3} (Q^2) - \Pi_{A,3} (Q^2) 
\right] \ \ .
\label{dim6}
\eeq

More interesting for our purposes here is the high energy portion. 
We  require that $\mu$ lie in the pQCD domain, 
and we further constrain it to obey $\mu \ll M_W$. The asymptotic
behavior of the vacuum polarization operator is then described by 
the operator product expansion, involving a series of
local operators ordered by increasing dimension. 
In the chiral limit the leading contribution
to the difference of vector and axial-vector correlators is a four-quark 
operator of dimension six\cite{svz,Lanin}, 
followed by a series of higher dimensional 
operators,  
\beq
(\Pi_{V,3} - \Pi_{A,3})(Q^2) \sim 
{2 \pi \langle \alpha_s {\cal O}^{(6)}_8 \rangle_{\mu} \over Q^6} + 
 {{\cal E}^{(8)}_\mu \over Q^8} + \dots \ \ .
\label{c2}
\eeq
Here ${\cal E}^{(8)}_\mu$ represents the combination of
local operators carrying dimension 
eight.

These have been discussed and 
partially calculated by Broadhurst and 
Generalis~\cite{bg}. For our purposes, it is not necessary to 
know their specific form, but only the fact of their existence.
Upon performing the integration over $Q^2$ at high energies, 
we find
\beq
{\cal M}_> (\mu) = {3 G_F \over 32 \sqrt{2}\pi^2 F_\pi^2} 
\left[ \ln \left( {M_W^2 \over \mu^2}\right) 2 \pi 
\langle \alpha_s {\cal O}^{(6)}_8 \rangle_{\mu} + 
{{\cal E}^{(8)}_\mu\over \mu^2} 
+ \dots \right] \ \ .
\label{dim7}
\eeq
In this expression we have dropped corrections of order 
$\mu^2 / M_W^2$. 

The full amplitude is then
\beq
{\cal M} \simeq {G_F \over 2 \sqrt{2}F_\pi^2}
\bigg[ \langle {\cal O}^{(6)}_1 \rangle_\mu^{\rm (c.o.)} \ 
+ \ { 3 \over 8 \pi} \ln \left( {M_W^2 \over \mu^2}
\right) \langle \alpha_s {\cal O}^{(6)}_8 \rangle_{\mu} \ +{3\over 16 
\pi^2}{{\cal E}^{(8)}_\mu \over \mu^2} + 
 \ldots \bigg] \ \ .
\label{sd11}
\eeq
The crucial features here are the presence of the dimension-eight 
operators in the short distance portion of the amplitude ${\cal M}$ 
and the fact that they appear divided by the scale $\mu^2$ 
instead of $M_W^2$.   They are {\it not} 
suppressed by inverse powers of $M_W$ because these operators 
appear in the vacuum polarization function at any $Q^2$ 
between $\mu^2$ and $\infty$.  From this calculation it is clear that  
these operators must be present in an OPE that describes the 
integrating-out of short distance physics.  
These operators have previously been missed in the usual treatment 
of the operator product expansion within the weak hamiltonian, 
although they have been properly included in 
the OPE for the vacuum polarization functions\cite{svz}. 
It will be clear from the work that we do below, where we find 
dimension-eight operators in the weak OPE, that they appear whenever
we separate physics above and below the scale $\mu$.  
Finally, despite our 
emphasis on the dimension-eight operators in this paper, operators 
of even higher dimension could play a role for sufficiently small 
values of $\mu$.  For example, the next term in Eq.~(\ref{sd11}) 
would be $3 {\cal E}^{(10)}_\mu/(32 \pi^2 \mu^4)$, 
where ${\cal E}^{(10)}_\mu$ represents the dimension-ten effect.  

\subsection{Estimated Size}

Here we give some numerical estimates of the size of the 
dimension-eight effects.  This can be done using experimental 
data since the vacuum polarization functions satisfy dispersion 
relations. The inputs to the dispersion integrals, {\it i.e.} 
the imaginary parts are known from experimental work on cross 
section measurements of $e^+ e^- \to$ hadrons and from the study 
of hadronic final states appearing in $\tau$ decay.  We have performed 
the required phenomenology in Ref.~\cite{dg}, and the reader is 
referred to that work for more detail. Here we use that reference 
to illustrate the size of various effects in the OPE.

First consider the sizes of the asymptotic elements in the vacuum 
polarization functions. Referring back to Eq.~(\ref{c2}), we can display 
the relative size of the coefficients of $Q^{-6}$ and $Q^{-8}$. 
We find this to be
\beq
{{\cal E}^{(8)}_{\mu \simeq 2~{\rm GeV}}  
\over 2 \pi \langle \alpha_s {\cal O}^{(6)}_8 
\rangle_{\mu \simeq 2~{\rm GeV}}} 
\simeq - 1.5 ~{\rm GeV}^2 \ \ .
\label{dim8}
\eeq
Thus the dimension-eight effect is quite relevant for the $\mu \sim 
1 \to 2$~GeV region.

Let us also look at the magnitude of the three terms in the OPE 
shown in Eq.~(\ref{sd11}). In the same order as displayed there 
(and in units of $10^{-7}$) we find
\beq
10^7~{\rm GeV}^{-2}~{\cal M}\  = \  \left\{ 
\begin{array}{ll}
- 0.12 \ - \ 3.84\  +\  0.64 + \dots & \phantom{xxxxx} 
(\mu = 1~{\rm GeV}) 
\nonumber  \\
- 0.28 \ - \ 3.49  \ + \ 0.30 + \dots & \phantom{xxxxx} 
(\mu = 1.5~{\rm GeV}) 
\nonumber  \\
- 0.44  \ - \ 3.24 \ + \ 0.17 + \dots & \phantom{xxxxx} 
(\mu = 2~{\rm GeV}) 
\nonumber  \\
- 0.89 \ - \  2.63 \ + \ 0.04 + \dots & \phantom{xxxxx} 
(\mu = 4~{\rm GeV}) \ \ .
\end{array}
\right.
\label{dim9}
\eeq

We see that at $\mu = 1$~GeV, the dimension-eight term is larger 
than the leading operator ${\cal O}^{(6)}_1$ in the OPE.~\footnote{We will see
that the appropriate comparison is with effect of ${\cal O}^{(6)}_1$ rather
than ${\cal O}^{(6)}_8$.} Even at 
$\mu =2$~GeV, it remains a significant size relative to this 
operator. However, at $\mu=4$~GeV it is clearly small enough to 
be neglected.

Through a great deal of effort, the short-distance perturbative 
structure of the weak interactions has been studied through 
two-loop order~\cite{Ciuchini,Buras}. 
The dimension-eight effects are large enough 
that their neglect would negate this effort, as we would be left with 
only crude evaluations, at least at values of $\mu$ which are 
presently used.
  
\subsection{Separation Scheme Dependence}

If one changes the separation scale, there is mixing between the 
operators of dimension six and dimension eight.  For example, if 
we use the scale $\mu_1$ instead of the scale $\mu$, the local 
operator at the new scale is related (in our perturbative 
treatment) to those at the old one by
\beq
\langle {\cal O}^{(6)}_1 \rangle_{\mu_1} = \langle {\cal O}^{(6)}_1\rangle_\mu 
+ {3\over 8\pi} 
\ln\left( {\mu_1^2 \over \mu^2}\right) \langle \alpha_s {\cal O}^{(6)}_8 \rangle
+ {3 {\cal E}^{(8)}_\mu\over 16\pi^2}
\left( {1\over \mu^2} - {1\over \mu_1^2} \right) \ \ .
\label{dim10}
\eeq
Thus portions of the dimension-eight effect will appear within the 
dimension-six operator evaluated at a given scale. 
 
There is also a dependence on the scheme by which one performs the 
separation of scales. In the above example, we used a sharp cutoff 
in the variable $Q^2$ as the method of dividing the low and high 
energy regions. This is certainly the most convenient method in 
the context of the present calculation, yet need not be the only 
possible method. Imagine a smoother cutoff $F(Q^2/\mu^2)$, with
$F(Q^2/\mu^2)\to 1$ for $Q^2 \ll \mu^2$ and $F(Q^2/\mu^2)\to 0$ for 
$Q^2 \gg \mu^2$.  We assume that this function is such that all 
the following integrals are well behaved.  Then let us define the 
vacuum matrix element of ${\cal O}^{(6)}_1$ in a so-called `$F$-scheme' as
\beq
\langle {\cal O}^{(6)}_1 \rangle_\mu^{\rm (F)} \equiv {3 \over 16 \pi^2}
\int_0^{\infty} dQ^2 ~Q^4 ~F\left( {Q^2\over\mu^2}\right)
\left[ \Pi_{V,3} (Q^2) - \Pi_{A,3} (Q^2)
\right] \ \
\label{dim11}
\eeq  
We also define the integrals
\beqa
c_F  & \equiv & - \ln\left({M_W^2\over \mu^2}\right) + 
\int_0^\infty {dQ^2 \over Q^2} \ {M_W^2\over Q^2 + M_W^2} 
~\left[ 1 - F\left( {Q^2\over\mu^2}\right) \right] 
\nonumber \\
{d_F \over \mu^2}  & \equiv & 
\int_0^\infty {dQ^2 \over Q^4} \ \left[ 1 - 
F\left( {Q^2\over\mu^2}\right) \right] \ \ .
\label{dim12}
\eeqa
We use these by inserting $1 = F + (1-F)$ into the full matrix
element, Eq.~(\ref{full}).  The first factor gives the ${\cal O}^{(6)}_1$ 
matrix element in the $F$ scheme, and the 
remaining integrals can be done.
In this scheme, the OPE for the amplitude ${\cal M}$ reads
\beq
{\cal M} \simeq {G_F \over 2 \sqrt{2}F_\pi^2}
\bigg[ \langle {\cal O}^{(6)}_1 \rangle_\mu^{\rm (F)} \
+ \ { 3 \over 8 \pi} 
\left[\ln \left( {M_W^2 \over \mu^2} \right) + c_F \right]
\langle \alpha_s {\cal O}^{(6)}_8 \rangle_{\mu} \ +{3 d_F \over 16
\pi^2}{{\cal E}^{(8)}_\mu\over \mu^2} +
 \ldots \bigg] \ \ .
\label{sd12}
\eeq
Therefore the matrix elements in the two schemes are related by
\beq
\langle {\cal O}^{(6)}_1 \rangle_\mu^{\rm (F)} =
\bigg[ \langle {\cal O}^{(6)}_1 \rangle_\mu^{\rm (c.o.)} \
- \ { 3c_F \over 8 \pi} 
\langle \alpha_s {\cal O}^{(6)}_8 \rangle_{\mu} \ +{3 (1 - d_F)\over 16
\pi^2}{{\cal E}^{(8)}_\mu\over \mu^2} +
 \ldots \bigg] \ \ .
\label{sd13}
\eeq
To fully specify the OPE and the matrix elements, one needs to clearly specify the
scheme for separating the scales. A lattice evaluation would involve different
combinations than does our initial sharp cutoff scheme.
We will return to this issue in the next section.

\subsection{Dimensional Regularization}

The presence of dimension-eight operators scaled by an 
inverse power of $\mu^2$ appears odd in the method of 
dimensional regularization since one expects only 
logarithms of $\mu$ in that scheme.  This is because in 
dimensional regularization one introduces an 
energy scale $\mu_{\rm d.r.}$ (`d.r.' denotes dimensional
regularization) in order to maintain the proper dimensions 
away from $d=4$. This scale appears only in the form
$\mu_{\rm d.r.}^{d-4}$ and as $d \to 4$, $\mu_{\rm d.r.}$
will appear only in logarithms. In this section we clarify 
this issue by means of explicit calculation.  
It will be seen that there exists a confusion between $\mu$ 
on the one hand and $\mu_{\rm d.r.}$ on the other.  The former 
is defined to be a separation scale (in the sense of effective 
field theory or of Wilson's 
OPE) whereas the latter has nothing to do with the 
separation of long and short distance. Indeed, dimensional 
regularization does not itself provide a mechanism for the 
separation of scales; {\it all} scales contribute to both the 
operator 
and the coefficient functions in such a scheme. 

We can evaluate the vacuum polarization functions as defined 
in $d$ dimensions,
\beqa
& & \mu_{\rm d.r.}^{d-4} i \int d^dx\ e^{i q \cdot x} 
\langle 0 |T\left( V^\mu_3  (x) V^\nu_3 (0) - A^\mu_3 (x) A^\nu_3 (0)\right) 
| 0 \rangle 
\nonumber \\
& & \phantom{xxxxx} = (q^\mu q^\nu - q^2 g^{\mu\nu} ) 
( \Pi_{V,3} - \Pi_{A,3} )(q^2) - q^\mu q^\nu \Pi_{A,3}^{(0)}(q^2) \ \ .
\label{r1}
\eeqa
>From this follows an expression for the 
vacuum matrix element of ${\cal O}^{(6)}_1$ in dimensional 
regularization, 
\beqa
\langle {\cal O}^{(6)}_1 \rangle_{\mu_{\rm d.r.}}^{\rm (d.r.)} &\equiv& 
\langle 0 |T\left( V^\mu_3  (0) V_{\mu,3}  (0) - 
A^\mu_3 (0) A_{\mu, 3} (0)\right) | 0 \rangle 
\nonumber \\
&=& { (d - 1) \mu_{\rm d.r.}^{4 -d} 
\over (4 \pi)^{d/2} \Gamma(d/2)} \int_0^\infty dQ^2 
\ ~Q^d  \left( \Pi_{V,3} - \Pi_{A,3} \right)(Q^2) \ \ .
\label{r2}
\eeqa
When $d < 4$, this expression is finite. 
A key point is that the integral continues to run over
all $Q^2$. It is not hard to 
relate this operator to the one found using a cutoff regularization.
To this end, we split the $Q^2$ integral into regions below and above 
$Q^2 =\mu^2$. Note that $\mu$ is not the same as $\mu_{\rm d.r.}$. 
For the part of the integration below separation scale $\mu^2$ 
the integral is finite for all dimensions, and we can take the 
limit $d \to 4$. This portion of the integration then 
reproduces exactly the cutoff version of the matrix element. We are 
left with the difference between the cutoff and dimensional 
regularization matrix elements.  It comes entirely from 
the high energy region, {\it i.e.} with $Q$ above the separation 
scale $\mu$ (see Ref.~\cite{dg} for details; here we use the NDR scheme 
for $\gamma_5$), 
\beqa
\lefteqn{ { (d - 1) \mu_{\rm d.r.}^{4 -d} \over (4 \pi)^{d/2} 
\Gamma(d/2)} \int_{\mu^2}^\infty dQ^2 \ Q^d
\left( \Pi_{V,3} - \Pi_{A,3} \right)(Q^2)}
\nonumber \\
& & \phantom{xxxxxx} = { (d - 1) \mu_{\rm d.r.}^{4 -d} 
\over (4 \pi)^{d/2} \Gamma(d/2)} \int_{\mu^2}^\infty dQ^2 \ Q^d 
\left[ {2\pi \langle \alpha_s O_8 \rangle \over Q^6} 
\left( 1 - {\epsilon \over 4} \right) 
+ {{\cal E}^{(8)}_\mu \over Q^8} + \dots \right] \nonumber \\
& & \phantom{xxxxxx} 
= {3 \over 8\pi} \langle \alpha_s O_8 \rangle \left[ {2\over 4-d} - \gamma 
+ \ln(4\pi) + \ln \left({\mu_{\rm d.r.}^2 \over \mu^2}\right) 
- {1 \over 6} \right]  + 
{3 \over 16 \pi^2} {{\cal E}^{(8)}_\mu\over \mu^2}  + \dots \ \ .
\label{dim13}
\eeqa
In ${\overline {\rm MS}}$ renormalization, the 
$2/(4-d) - \gamma + \ln(4\pi)$ terms are removed.  
Completing the calculation one finds
\beq
\langle {\cal O}^{(6)}_1 
\rangle_{\mu_{\rm d.r.}}^{\rm {\overline {\rm (MS)}}}
= \langle {\cal O}^{(6)}_1 \rangle_\mu^{\rm (c.o.)}
+ {3 \alpha_s \over 8 \pi} \left[ \ln\left({\mu_{\rm d.r.}^2 \over 
\mu^2}\right) - {1\over 6}\right] \langle {\cal O}^{(6)}_8 \rangle_\mu 
+ {3\over 16 \pi^2}{{\cal E}^{(8)}_\mu \over \mu^2} \ \ .
\label{va37}
\eeq
Thus, all of the dimension-eight operator is shifted into the 
${\overline {\rm MS}}$ definition of the dimension-six operator. 
This is seen to be consistent: 
\begin{enumerate}
\item When one performs a separation of scales, one has the need 
for dimension-eight operators in the OPE scaled by
$1 / \mu^2$. 
\item When one defines instead the OPE using dimensional
regularization, one cannot get effects proportional to 
$1 / \mu^2_{\rm d.r.}$, but the same effect appears contained 
within the dimension-six operator matrix element. Inspection
of Eq.~(\ref{sd11}) shows that one obtains the same 
overall matrix element\footnote{Since we are treating the 
dimension-six coefficients at leading-log order, we can 
ignore the nonlogarithic dimension-six portion of Eq.~(\ref{va37}). 
To include it only requires an inclusion of the nonlogarithmic 
terms in the coefficient function.}.
\end{enumerate}

It is important to recognize that dimensional regularization is not 
a true separation of scales for the OPE in the original sense 
meant by Wilson.~\cite{Wilson1,Wilson2,Wilson3,Wilson4}  
The dimensionally regularized
matrix element contains effects from {\em all} scales, including 
finite but sizeable contributions from short distances. Thus 
any use of dimensional regularization for the coefficient functions 
must be accompanied by an evaluation of the matrix element that 
covers all scales.

The problem with this situation is that, in present 
practice, the matrix elements are always calculated with some form
 of a cutoff but the coefficient functions are calculated dimensionally. 
This is an inconsistent procedure. To relate operators in a cutoff scheme
to dimensional ones, a dimension-eight effect needs to be included, as 
in Eq.~(\ref{va37}). We will return to a more complete 
discussion of this point later. 
   
\section{Dimension-eight Operators in the Standard Model} 

In this section we calculate the dimension-eight operators (and their
Wilson coefficients) which are relevant for the Standard Model 
at one loop. We present these first employing a cutoff to provide a 
separation of scales, then return to a discussion of how to 
use these results in the context of dimensional regularization.

\subsection{Defining a Cutoff Procedure}

The construction of the OPE is performed by comparing a 
calculation performed both in the full Standard Model and 
within the effective theory. The coefficients of the effective 
theory are adjusted such that the results of the two 
are identical to a given order. For this purpose,
calculations using free quarks and gluons are simplest, and
are sufficient to identify the operators and their coefficients.

We perform this calculation to one-loop order.
For the full Standard Model amplitude,
one calculates up to ${\cal O}(\alpha_s)$ in the usual way.
For the effective theory one takes all matrix elements at
tree level, except for those of the leading 
operator,
\beq
{\cal Q}^{(6)}_{2} \equiv \bar{u} \Gamma_{\mu}^{L} s ~
\bar{d} \Gamma^\mu_L u \ \ ,
\label{q2}
\eeq
which must be calculated at one-loop level in order to
properly include the ${\cal O}(\alpha_s)$ effects. In
both the full and effective theories one adopts the same 
external states and kinematics.  It is important, however, to 
employ a distinct four-momentum for each of the external states.

We use a cutoff to regularize the matrix element in the following way.
Consider the matrix element of a current product at different
spacetime points, 
\beq
{\cal M}_{A\to B} = \langle B|T(J_\mu^{\rm ch} (x)
J_{\rm ch}^{\dagger\mu} (0) )|A\rangle  \ \ .
\eeq
where $J_{\rm ch}^\mu$ is the hadronic charged weak current.
Since the current is a color singlet, ${\cal M}_{A\to B}$ is 
invariant under QCD
gauge transformations for any value of $x$. In order to define an
operator matrix element from this, we fourier transform to momentum
space, rotate to euclidean momentum and apply a cut-off on the
euclidean momentum such that $Q^2 \le \mu^2$, 
\beq
\langle B|O|A\rangle_\mu = \int {d^4Q \over (2\pi)^4} \ 
\Theta (\mu^2-Q^2) \int
d^4x ~ e^{iQx}
\langle B|T(J_\mu (x) J^\mu (0) )|A\rangle  \ \ .
\eeq
This procedure puts the low-$Q^2$ components into the matrix element
and leaves the high-$Q^2$ portion to be accounted for in the OPE.
The analysis of the high-$Q^2$ portion is then accomplished in
the same way as in QCD sum rules via the specification of
the momentum flowing in the currents. 
In practice, this amounts to the following recipe. 
While calculating 
insertions in one-loop diagrams, we imagine the two currents to be
connected by a $W$-like boson with virtual four-momentum $k$. We
route the virtual momenta in the loop according to this prescription,
choosing $k$ as the loop-integration variable and regularizing
the theory by means of a sharp cutoff $\mu^2$ on the euclidean
squared-momentum $k_E^2$.  The result obtained this way is UV finite,
and we {\it define} it as the matrix element at the scale $\mu$.
 
Given this separation of scales, the OPE operators can be calculated in
one of two equivalent ways.
\begin{enumerate}
\item We first calculate a matrix element in the full theory, which will 
be finite and independent of $\mu$. The {\em low energy} 
radiative corrections
to the operator ${\cal Q}_2^{(6)}$ below the scale $\mu$ are 
then calculated in the
effective theory. While the infrared portions of these matrix elements 
will be the same, a comparison of the two reveals that specific local
operators need to be added to the effective theory in order to reproduce 
the results of the full theory. These are the new operators in the OPE at 
one loop. In this method, dimension-eight operators are needed because 
radiative corrections of the matrix
elements in the effective theory contain $1/\mu^2$ effects,
which do not occur in the full theory, and hence must be corrected for
by dimension-eight operators.
\item The same result can be found in the {\em high energy} portion of 
the calculation. In this method, the portion of the full theory 
that occurs above the scale $\mu$ is considered. This portion is equivalent
to a set of local operators and the calculation readily reveals their 
coefficients.~\cite{nsvz} In this case, it is seen that 
dimension-eight operators are a real contribution to
the matrix element, and the factor of $1/\mu^2$ arises simply
as the lower end of the region of momentum being considered. It is this 
method which was used in the first portion of this paper.  
\end{enumerate}
These two methods are equivalent because the net physics is 
independent of $\mu$.

\subsection{Results}
We summarize our calculation of dimension-eight 
contributions to both the current-current operators 
and the QCD penguin vertex.  {\it Throughout we work 
in the chiral limit of $m_u = m_d = m_s = 0$}. This is an
appropriate and useful limit because it captures the leading chiral 
contribution to kaon matrix elements. The leading weak chiral lagrangian 
that contains factors of the quark masses can be diagonalized 
away by a chiral rotation. This implies that the effects of quark masses 
are suppressed by one order in the chiral expansion.     

\vspace{0.3in}

\begin{figure}
\vskip .1cm
\hskip 2.0cm
\psfig{figure=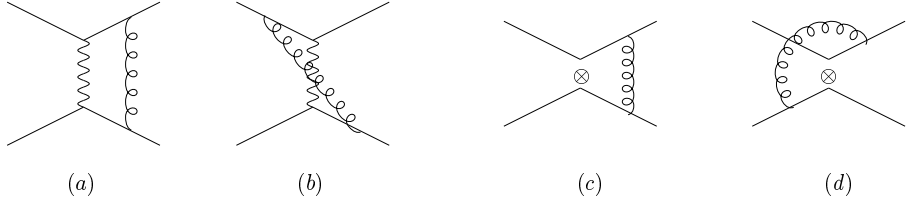,height=1.in}
\caption{QCD corrections to the box: full theory (a)-(b), 
effective theory (c)-(d).\hfill 
\label{fig:dim8fig2}}
\end{figure}

{\it Box Diagrams}: First, we consider the 
current-current sector of the $\Delta S = 1$ 
four-quark sector which arises from the 
box diagram.  We depict in Fig.~\ref{fig:dim8fig2} 
both the `full' and `effective' descriptions for 
one of the four possible box-diagram contributions.  
Virtual quarks which appear within a box loop are given a (very 
small) common mass `$m$' to regularize infrared behavior.  

To zeroth order in QCD 
the current-current hamiltonian is expressible entirely 
in terms of the dimension-six operator ${\cal Q}_2^{(6)}$ 
({\it cf} Eq.~(\ref{q2})), 
\beq
{\cal H}_{\rm eff}\bigg|_{\rm No~QCD} 
= {G_F \over \sqrt{2}} V_{us}V_{ud}^* ~
{\cal Q}_2^{(6)}  \ \ . 
\label{curcura}
\eeq
The inclusion of QCD to first order yields the 
familiar dimension-six component, which we express in 
the `color-basis' as 
\beq
{\cal H}_{\rm (curr-curr)}^{(6)}   
= {G_F \over \sqrt{2}} V_{us}V_{ud}^* 
\left[ {\cal C}_2^{(6)} {\cal Q}_2^{(6)} + 
{\cal C}_C^{(6)} {\cal Q}_C^{(6)} \right] \ \ ,
\label{curcurb}
\eeq
where 
\beq
{\cal Q}_C^{(6)} = {\bar d} \Gamma^\mu_a u ~ {\bar u} 
\Gamma_\mu^a s 
\label{opcolor}
\eeq
and 
\beq
{\cal C}_2^{(6)}(\mu) = 1 + {\cal O}(\alpha_s^2) \ , \qquad 
{\cal C}_C^{(6)}(\mu) = - {3\over 2} {\alpha_s \over \pi} 
\ln\left({M_W^2 \over \mu^2}\right) \ \ .
\label{wilsona}
\eeq 

\begin{figure}
\vskip .1cm
\hskip 2.0cm
\psfig{figure=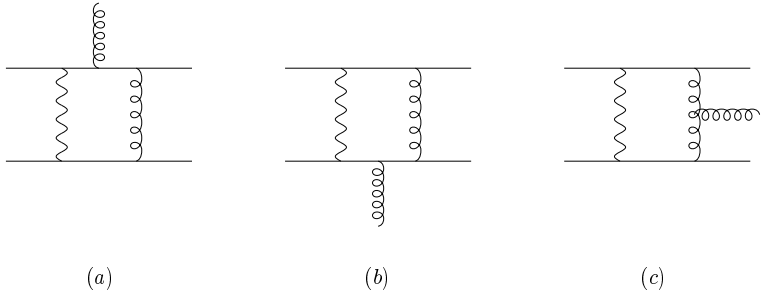,height=1.5in}
\caption{Gluon radiation from the box diagram.\hfill 
\label{fig:dim8fig4}}
\end{figure}

The set of dimension-eight current-current operators is 
constructed from four-quark products, covariant derivatives 
and gluon field-strength tensors.  The gluon field-strength 
tensors we shall employ in our analysis are 
\beq
F_{\mu\nu}^a = \partial_\mu G^a_\nu - \partial_\nu G^a_\mu 
- g_3 f_{abc} G_\mu^b G_\nu^c  \ , \qquad 
{\tilde F}_{\mu\nu}^a = {1 \over 2} \epsilon_{\mu\nu\alpha\beta} 
F_a^{\alpha\beta} \ \ .
\label{str}
\eeq
As such, there will be contributions from gluon emission graphs (see 
Fig.~\ref{fig:dim8fig4}).  Also, in constructing the 
dimension-eight operator basis, we have used the quark equation 
of motion and current conservation in the chiral limit,  
\beq
\gamma \cdot {\cal D} ~q = 0 \ , \qquad {\cal D}_\mu \left( 
{\bar q}_1 \Gamma^\mu q_2 \right) = 0 \ \ .
\label{chrelns}
\eeq
The list of gauge invariant dimension-eight operators is 
then\footnote{It might appear that the list in Eq.~(\ref{boxops}) 
is missing another possible dimension-eight operator, namely 
$g_3 f_{abc} F^{\mu\nu,b} {\bar u} \Gamma_\nu^a s~ 
{\bar d} \Gamma_\mu^c u$.  However, one can show that in the 
chiral limt, this is expressible in terms of  
${\cal Q}_5^{(8)}$ and ${\cal Q}_6^{(8)}$.}   
\beqa
{\cal Q}_1^{(8)} &=& 
{\bar u} \stackrel{\leftarrow}{{\cal D}_\mu} 
\stackrel{\leftarrow}{{\cal D}^\mu} \Gamma_\nu^a s~ 
{\bar d} \Gamma_a^\nu u + 
 {\bar u} \Gamma_\nu^a {\cal D}_\mu {\cal D}^\mu  s~ 
{\bar d} \Gamma_a^\nu u   
 + {\bar u} \Gamma_\nu^a s~ 
{\bar d} \stackrel{\leftarrow}{{\cal D}_\mu} 
\stackrel{\leftarrow}{{\cal D}^\mu} \Gamma_a^\nu u   
 + {\bar u} \Gamma_\nu^a s~ {\bar d} 
\Gamma_a^\nu {\cal D}_\mu {\cal D}^\mu u   
\nonumber \\
{\cal Q}_2^{(8)} &=& 
{\bar u} \Gamma_\nu^a {\cal D}_\mu s~ 
{\bar d} \Gamma_a^\nu {\cal D}^\mu u   + 
{\bar u} \stackrel{\leftarrow}{{\cal D}^\mu} \Gamma_\nu^a s~ 
{\bar d} \stackrel{\leftarrow}{{\cal D}_\mu} \Gamma^\nu_a u   
\nonumber \\
{\cal Q}_3^{(8)} &=& 
{\bar u} \stackrel{\leftarrow}{{\cal D}_\mu} 
\Gamma_\nu^a s~ {\bar d} \Gamma_a^\nu {\cal D}^\mu u   + 
{\bar u} \Gamma_\nu^a {\cal D}_\mu s~ 
{\bar d} \stackrel{\leftarrow}{{\cal D}^\mu} \Gamma_a^\nu u   
\nonumber \\
{\cal Q}_4^{(8)} &=& 
 {\bar u} \stackrel{\leftarrow}{{\cal D}_\mu} 
\stackrel{\leftarrow}{{\cal D}_\nu} \Gamma^\nu_a s~ 
{\bar d} \Gamma_a^\mu u   
 + {\bar u} \Gamma^\nu_a {\cal D}_\nu {\cal D}_\mu  s~ 
{\bar d} \Gamma_a^\mu u   
 + {\bar u} \Gamma^\nu_a s~ 
{\bar d} \stackrel{\leftarrow}{{\cal D}_\nu} 
\stackrel{\leftarrow}{{\cal D}_\mu} \Gamma_a^\mu u   
 + {\bar u} \Gamma^\nu_a s~ {\bar d} 
\Gamma_a^\mu {\cal D}_\mu {\cal D}_\nu u   
\nonumber \\
{\cal Q}_5^{(8)} &=& 
{\bar u} \stackrel{\leftarrow}{{\cal D}_\mu} \Gamma_a^\nu s
~{\bar d} \stackrel{\leftarrow}{{\cal D}_\nu} 
\Gamma_a^\mu u + {\bar u} \Gamma_\nu^a {\cal D}_\mu s~ 
{\bar d} \Gamma_a^\mu {\cal D}^\nu u   
\nonumber \\
{\cal Q}_6^{(8)} &=& 
 {\bar u} \stackrel{\leftarrow}{{\cal D}_\mu} \Gamma_a^\nu s 
~{\bar d} \Gamma_a^\mu {\cal D}_\nu u   
+ {\bar u} \Gamma^\nu_a {\cal D}_\mu s~ 
{\bar d} \stackrel{\leftarrow}{{\cal D}_\nu} \Gamma_a^\mu u  
\nonumber \\
{\cal Q}_7^{(8)} &=& g_3 
\delta^{ab} {\tilde F}^{\mu\nu,b} \left[ 
{\bar u} \Gamma_\mu^a s~ {\bar d} \Gamma_\nu u - 
{\bar u} \Gamma_\mu s~ {\bar d} \Gamma_\nu^a u \right] \ \ ,
\label{boxops}
\eeqa
where $\stackrel{\leftarrow}{{\cal D}_\mu}$ denotes a left-acting 
operation and we define the convenient notation 
\beq
\Gamma^a_\mu \equiv {\lambda^a \over 2} \Gamma^L_\mu \equiv 
{\lambda^a \over 2} \gamma_\mu (1 + \gamma_5) \ \ .
\label{denote1}
\eeq

We find the Wilson coefficients corresponding to the above 
dimension-eight operators to have the form
\beq
{\cal C}_i^{(8)} \ = \ {\alpha_s \over 4 \pi} \cdot {1 \over \mu^2} 
\cdot \eta_i^{(8)} \ \ ,
\label{wilcoeff}
\eeq
where the $\{ \eta_i^{(8)} \}$ coefficients are: 

\beq
\begin{array}{c|c|c|c|c|c|c|c|}
\eta_1^{(8)} & \eta_2^{(8)} & \eta_3^{(8)} & \eta_4^{(8)} & 
\eta_5^{(8)} & \eta_6^{(8)} & \eta_7^{(8)}  \\ 
\hline 
5/3 & 22/3 & 8/3 & - 1/3
& 16/3 & 14/3 & 1/3 
\end{array} \ \ .
\eeq

\vspace{0.3in}

\begin{figure}
\vskip .1cm
\hskip 2.0cm
\psfig{figure=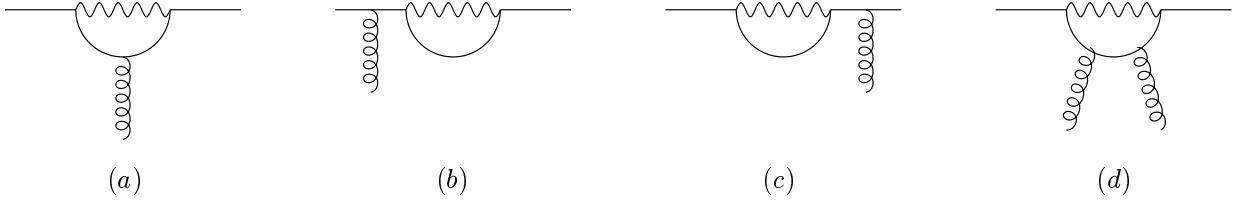,height=1.in}
\caption{QCD penguin vertex in the full theory.\hfill 
\label{fig:dim8fig3}}
\end{figure}

{\it QCD Penguin Vertex}:  In the chiral limit, 
the general form for the QCD penguin effective vertex is 
\beq
{\cal H}_{\rm eff}^{\rm (QCD-pgn)} = {G_F \over \sqrt{2}} \cdot 
V_{us}V_{ud}^* \left[ {\cal C}_{\rm P}^{(6)} 
{\cal O}_{\rm P}^{(6)} +  
\sum_{i = 1}^3 ~{\cal C}_{{\rm P}i}^{(8)} 
{\cal O}_{{\rm P}i}^{(8)} + \dots \right] \ \ .
\label{penver}
\eeq
Here, we employ a two-generation approximation (no dependence on 
the virtual top quark in the penguin loop). 
The penguin vertex as it appears in the `full' theory 
is displayed in Fig.~\ref{fig:dim8fig3}.  Observe the presence 
of self-energy graphs in Figs.~\ref{fig:dim8fig3}(b),(c).

The $d=6$ QCD-penguin operator and its Wilson coefficient 
are given (in the chiral limit) by 
\beq
{\cal O}_{\rm P}^{(6)} = \bar{d} \Gamma_\nu^a s ~
{\cal D}_\mu F^{\mu\nu}_a \ , \qquad 
{\cal C}_{\rm P}^{(6)} = {g_3 \over (4\pi)^2} \left[ 
-{4 \over 3} \ln {\mu^2 + m_c^2 \over \mu^2} + 2 g_c 
- 2 g_c^2 + {2\over 9} g_c^3 \right] \ \ , 
\label{pensix}
\eeq
where $g_3$ is the QCD coupling constant 
and $g_c$ is the dimensionless quantity 
\beq
g_c \equiv {m_c^2 \over \mu^2 + m_c^2} \ \ .
\label{pendefn}
\eeq

Proceding next to the dimension-eight component, there 
are two classes of local, gauge-invariant operators: 
with an $s \to d$ quark bilinear and one field-strength 
tensor (${\cal O}_{{\rm P}1}^{(8)}$) and 
with an $s \to d$ quark bilinear and two 
field-strength tensors ($\{ {\cal O}_{{\rm P}i}^{(8)} \} \ (i = 2,3)$).  
The latter correspond to two-gluon emission as in 
Fig.~\ref{fig:dim8fig3}(d).   Thus we find 
\beqa
{\cal O}_{{\rm P}1}^{(8)} &=&  \bar{d} \Gamma_\nu^a s ~ 
{\cal D}^\alpha {\cal D}_\alpha {\cal D}_\mu F^{\mu\nu}_a 
\ \ , \nonumber \\
{\cal O}_{{\rm P}2}^{(8)} &=& i \bar{d} \left[ {\lambda^a \over 2} , 
{\lambda^b \over 2} \right] \Gamma^\mu_L s ~ 
F_b^{\alpha\beta} {\cal D}_\mu F_{\alpha\beta}^a 
\ \ , 
\nonumber \\
{\cal O}_{{\rm P}3}^{(8)} &=& \bar{d} \left[ {\lambda^a \over 2} , 
{\lambda^b \over 2} \right]_+ \Gamma^\mu_L s ~ 
F_a^{\alpha\beta} {\cal D}_\alpha {\tilde F}_{\beta\mu}^b \ \ , 
\label{pen8op}
\eeqa
along with the $d=8$ Wilson coefficients, 
\beqa
{\cal C}_{{\rm P}1}^{(8)}(\mu) &=& {g_3 \over (4\pi)^2} 
{1 \over \mu^2 + m_c^2} \left[ {1 \over 3} {m_c^2 \over
\mu^2} - {2 \over 3} g_c + {4 \over 3} g_c^2 - {2\over 3} g_c^3 
+ {1\over 15} g_c^4 \right] \ \ ,
\nonumber \\
{\cal C}_{{\rm P}2}^{(8)}(\mu) &=& {\alpha_s \over 3\pi} 
{1 \over \mu^2 + m_c^2} \left[ - {4 \over 3} {m_c^2 \over \mu^2} 
- 3 g_c + {10 \over 3} g_c^2 -  {13 \over 6} g_c^3 
+ {2\over 5} g_c^4 \right] \ \ , 
\nonumber \\
{\cal C}_{{\rm P}3}^{(8)}(\mu) &=& {\alpha_s \over 3\pi} 
{1 \over \mu^2 + m_c^2} \left[ {4 \over 3} {m_c^2 \over \mu^2} 
+ 2 g_c - {4 \over 3} g_c^2 -  
{1\over 3} g_c^3 \right] \ \ . 
\label{pen6coeffb}
\eeqa

\subsection{Conversion to dimensional regularization}

As we have stated, the dimension-eight operators
and Wilson coefficients of the previous section
are defined in terms of a particular momentum cutoff procedure.
It is essential to understand how these relate to those
obtained via other possible calculational approaches, most notably
dimensional regularization and ${\overline{\rm MS}}$ renormalization.

In dimensional regularization, there are no $1/\mu^2$ effects. 
Therefore all dimension-eight effects must be transfered from
existing as explicit operators in the OPE to being contained within 
some ${\overline{\rm MS}}$ matrix element. This was previously illustrated 
in our sample calculation Sect.~II as embodied in Eq.~(\ref{va37}). 
In the present case we are looking at radiative corrections to the 
operator ${\cal Q}_2^{(6)}$ and therefore at this
order all the dimension-eight effects will be absorbed into the 
${\overline{\rm MS}}$ matrix element of ${\cal Q}_2^{(6)}$. It has 
long been realized that to convert to ${\overline{\rm MS}}$ 
from some form of cutoff in the evaluation of matrix elements, a
mixing of dimension-six operators is needed. What is new in the present 
work is the realization that dimension-eight physics is also needed. 
In terms of the coefficients $\{ {\cal C}_i^{(8)}\}$ calculated above, 
the relation is
\beq
\langle {\cal Q}_2^{(6)}\rangle ^{{\overline{\rm MS}}}_\mu 
= \langle {\cal Q}_2^{(6)}\rangle^{c.o}_\mu + \sum_{i} d_i 
\langle {\cal Q}_i^{(6)}\rangle_\mu 
+ \sum_{i}  {\cal C}^{(8)}_i \langle O_i^{(8)}\rangle \ \ .
\eeq 
Here $d_i$ are mixing coefficients for dimension six which we do not 
calculate in this paper. However, the key new feature here is that
the matrix elements of the dimension-eight operators must be added 
directly to that of dimension six in order to form the 
${\overline{\rm MS}}$ matrix element.  Other ${\overline{\rm MS}}$ 
operators will also have similar relations involving dimension-eight 
effects when converting from cutoff schemes.  We discuss this point 
in the Appendix.

\section{Conclusion}

We have uncovered a basic problem with existing calculations 
of weak amplitudes.  The coefficients in the OPE are calculated 
using dimensional regularization and hence do not include 
dimension-eight operators. However, all matrix elements
are calculated with some variation of a cutoff and hence also 
do not contain the effects of dimension-eight operators. When 
one connects the matrix elements to those in the 
${\overline {\rm MS}}$ scheme, one needs to consider {\it both} 
dimension-six and dimension-eight operators as our example shows 
in Eq.~(\ref{va37}). However in practice this is not done, and 
the result is then inconsistent. 

Lattice evaluations of operators\cite{lattice}
 rely on the finite lattice spacing to
remove all short distance physics. This is a true cutoff, as the 
effects of short distance physics is simply not present in the 
simulation. Most current lattice calculations identify the inverse
lattice spacing with the scale $\mu$ that defines the matrix element, 
although there are alternative possibilities (see item 1 below for 
an example). Therefore, to convert from lattice regularization 
at any fixed lattice spacing to a dimensional scheme
requires the addition of both dimension-six 
and dimension-eight operators. This can be done as part of the 
`improvement' procedure\cite{symanzik} 
which attempts to correct for the effects
of lattice artifacts, including the effects of finite lattice spacing. 
On the lattice the effects of higher dimension operators can be more 
severe, as the lack of chiral invariance can generate a dimension 
seven operator\cite{chris}, whose effect also needs to be corrected for.
The present state of the art does not yet include the effects
of dimension eight operators, but in the future it should be possible 
to correct also also for these.

Quark model (and also large N$_c$) matrix elements are in a far more 
difficult situation\cite{quarkmodel}. 
In these cases, low energy models are postulated,
and the models only make sense below 1~GeV or so.  These models 
are treated with a cutoff of order 1~GeV, and cannot contain the required 
short-distance physics. While lattice methods are correctible to account
for dimension-eight effects, the same appears doubtful in quark model
methods. We then must conclude that there are very large intrinsic uncertainties
associated with these methods.

\vspace{0.4cm}

There are several strategies for overcoming the basic inconsistency 
we have revealed:
\begin{enumerate}
\item In principle, the best way is avoid the need for a cutoff in 
the matrix element calculation - to have a method which is sensitive
to all scales. This can then directly yield ${\overline{\rm MS}}$ 
matrix elements.  One possible procedure is the lattice analysis of 
the kaon $B$-parameter in Ref.~\cite{sharpe}, where the
scale $\mu_{\rm d.r.}$ is held fixed while the lattice spacing is
varied. The various values of the lattice spacing are then
used to perform an extrapolation to the continuoum limit, 
at fixed $\mu_{\rm d.r.}$.  Although this was done to remove lattice 
artifacts, it also has the effect of removing the need for 
dimension-eight effects (since the separation scale in the matrix 
element is sent to infinity).  A second procedure is the evaluation 
that we provided in the first half of this paper, based on sum 
rule techniques where one knows the vacuum polarization functions 
at all scales~\cite{dg,others}.  In this method there is no obstacle to 
a full evaluation, because we can take the separation scale to 
infinity\footnote{This will be discussed more fully in a separate 
work by us.}. See also Ref.~\cite{peris} for a calculation sensitive to 
all distance scales. In general, the lattice seems most suited for a 
systematic program of extrapolating matrix elements to very large 
separation scales. We caution, however, that this has not yet 
been done in all cases of interest.

\item A less ambitious but still valuable strategy would be to 
work at values of separation scale $\mu$ large enough that 
the problems of higher dimensions are numerically insignificant.  
Stated in a different way, at large enough $\mu$
the matrix element evaluation will already contain
all the needed ingredients, and residual dimension-eight effects that 
are missing will be small enough. This option is likely 
only available for lattice regularization, where it is tied to the 
ability to decrease the lattice spacing. Quark model methods make 
sense only at low energy and cannot be extended to larger $\mu$. 
The question in this case is how large $\mu$ must be for a result 
of a given accuracy. We have addressed this for our model 
hamiltionian above with the result that $\mu \sim 3-4$~GeV appears 
to be required.  If the lattice is used to give estimates of matrix 
element of the dimension eight operators calculated in Sect.~III, 
these results could be used to provide a second estimate of the 
relative importance of these terms. An advantage of this method 
is that one does not need to be highly accurate or highly consistent. 
If the dimension-eight effects are small at a given $\mu$, then we do 
not need to match them to the rest of the calculation. We instead use 
the estimated size to produce an error bar due to this effect. If the 
error bar is small enough, we need need not consider dimension-eight 
operators further.

\item Alternatively, one might adopt a cut-off scheme for both the 
matrix elements and the OPE coefficients. This would bring 
dimension-eight operators into the OPE, 
and their matrix elements would also need to be computed. However, this 
option appears difficult to carry out to the accuracy that we desire. At 
one loop, a cut-off scheme is only moderately difficult. One needs to be careful
to preserve gauge invariance and other symmetres. One also needs to find
a scheme that is equally useful for the perturbative calculation of the 
coefficients and for the calculation of the matrix elements if one is going 
to match them consistently. This latter requirement is likely difficult. However,
it will be extremely difficult to implement a cut-off scheme at two loops. 
The presence of a dimensionful parameter upsets the normal power counting, 
and requires great care in properly defining the operators beyond one loop.
In addition, the separation of nested loops in a cutoff scheme is 
subtle and can potentially lead to troubles with gauge invariance and other 
symmetries. 

\item Yet another option would be to continue to follow 
the most standard practice (of using dimensional regularization 
for the Wilson coefficients and a cutoff procedure for the 
operator matrix elements), but to correct the matrix elements 
to include the short distance effects from
dimension-eight operators. In this case one would 
calculate both dimension-six and dimension-eight operators in 
some cutoff scheme, then add the contributions together with 
the right coefficient to form the 
dimensionally regularized matrix element.  This is what is 
done earlier in Eq.~(\ref{va37}).  This allows one to use the 
extensive work that has been performed calculating
the OPE coefficients. It remians to be seen if this procedure can be
successfully carried out in all cases of interest.

\end{enumerate}

Much of the existing work in the field is done at low values 
of $\mu$. Various quark model and large-N$_c$ methods use $\mu\sim 0.5 
\to 1$~GeV, at which scales these effects are apparently extremely 
important. These calculations must be considered to contain enormous 
uncertanties, at least until further work is done. Lattice 
calculations are typically carried out with $\mu = 2$~GeV, although 
there is recent progress at working at higher $\mu$. At this 
scale, dimension-eight contributions appear to still be larger 
than other effects which are included, such as scheme dependence 
and two-loop evaluation of the coefficient functions. 
We look forward to future work that allows us to eliminate or 
reduce the undertainties that come from the presence of 
dimension-eight effects.

\acknowledgements
This work was supported in part by the National
Science Foundation.  One of us (V.C.) acknowledges 
support from the Foundation A. Della Riccia.
Two of us (V.C. and J.D.) thank 
the CERN Theory Division, where part of the work described 
in this paper was carried out, for its hospitality. 
We are grateful to C. Sachrajda for useful conversations.

\eject

\appendix

\section{On the Matching of Cutoff and Dimensional 
Regularization Procedures} 

In Sect.~III, we utilized the dimension-six local operator 
${\cal Q}_2^{(6)}$ in our discussion of higher dimension 
effects in the Standard Model.  Of course, 
a similar analysis can be performed for other operators 
${\cal Q}_i^{(6)}$ appearing in the weak hamiltonian. 
Here, we consider briefly the issue of perturbative matching between
cutoff and dimensional regularization approaches in a general 
situation.

Let us first establish the notation. We let ${\cal Q}_i^{\rm (6)} (\mu)$
represent the $d=6$ operators normalized in any cutoff scheme 
at the  scale $\mu$ and ${\cal Q}_i^{(6){\overline {\rm MS}}} 
(\mu_{\overline{\rm MS}})$ represent the corresponding 
dimension-six operator in the $\overline{\rm MS}$ normalization.  
Further, we let $S_i^{(d)}$ denote a tree-level matrix element 
($S_i^{(d)} \equiv \langle {\cal Q}_i^{(d)} 
\rangle^{\rm (tree)}$) and consider matrix elements taken between 
states with generic momentum $p$ and quark mass $m_q$.  

Then, calculating in a cutoff scheme, and keeping terms up to 
dimension eight will give generally 
\beq
\langle {\cal Q}_i^{(6)} \rangle^{\rm (c.o.)}_\mu = \left[ 
\delta_{ij} + \alpha_s \gamma_{ij} \ln\left({\mu^2 \over p^2}\right) 
+ \alpha_s f_{ij}^{\rm c.o.} \right] S_j^{(6)} + 
\alpha_s \left[ c_{ij} + {1 \over \mu^2} 
\tilde{\gamma}_{ij} \right] S_j^{(8)} \ \ . 
\label{dr4}
\eeq
Here $\gamma_{ij}$ is the usual dimension-six anomalous 
dimension matrix. 
The coefficients $c_{ij}$ scale as $p^{-2}$ or $m_q^{-2}$ and 
therefore are sensitive to the infrared (IR) behavior of the matrix 
element, while $\tilde{\gamma}_{ij}$ is associated with the behavior 
at the upper end of the integration domain. 
$f_{ij}^{\rm c.o.}$ are finite terms depending on the specific scheme 
adopted for separating the scales. 
For finite $\mu$ the matrix element $\langle {\cal Q}_i^{(6)}
\rangle^{\rm (c.o.)}_\mu$ will itself be finite and could be used as
the definition of the operator matrix element at scale $\mu$. 
Of course, the operators defined in this way differ from the 
$\overline{\rm MS}$ ones by a finite normalization. 

We now present the connection to the operators in the 
${\overline {\rm MS}}$ scheme. This requires an appropriate 
matching in which dimension-eight operators will appear.  
An example of this is provided by Eq.~(\ref{va37}) in the study 
of the LR hamiltonian.  Consider then a matrix element 
of the $d=6$ operator ${\cal Q}_i^{(6)}$ 
taken between four-quark states. Expressed schematically, 
the calculation as performed in dimensional regularization 
(`d.r.') gives 
\beq
\langle {\cal Q}_i^{(6)} \rangle^{\rm (d.r.)}_{\mu_{\rm d.r.}}
 = \left[ 
\delta_{ij} + \alpha_s \gamma_{ij} \left( 
\ln{\mu_{\rm d.r.}^2 \over p^2} 
+ {1 \over {\hat \epsilon}} \right) + \alpha_s f_{ij}^{\rm d.r.} 
\right] S_j^{(6)} + \alpha_s c_{ij} S_j^{(8)} \ \ ,
\label{dr1}
\eeq
with 
\beq
{1 \over {\hat \epsilon}} \equiv {2 \over \epsilon} - \gamma 
+ \ln(4 \pi) \ \ .
\label{e-hat}
\eeq
The infrared coefficients $c_{ij}$ are the same as in the case of the
cutoff scheme. In this case, however, the terms proportional to 
$\tilde{\gamma}_{ij}$ disappear because in dimensional regularization 
the integration runs over all scales.  Introducing an 
$\overline{\rm MS}$-subtracted operator,  
\beq
\langle {\cal Q}_i^{(6)}\rangle^{\overline{\rm (MS)}} 
\equiv \left[ \delta_{ij} - 
\alpha_s \gamma_{ij} {1 \over {\hat \epsilon}} \right] 
\langle {\cal Q}_j^{(6)} \rangle^{\rm (d.r.)} \ \ , 
\label{dr2}
\eeq
we have 
\beq
\langle {\cal Q}_i^{(6)}
\rangle^{\overline{\rm (MS)}}_{\mu_{\overline{\rm MS}}} 
= \left[ 
\delta_{ij} + \alpha_s \gamma_{ij} 
\ln \left({\mu_{\overline{\rm MS}}^2 \over p^2} \right)
+ \alpha_s f_{ij}^{\rm d.r.} \right] S_j^{(6)} + 
\alpha_s c_{ij} S_j^{(8)} \ \ .
\label{dr3}
\eeq

The general form of the above mentioned connection at order $\alpha_s$ 
is given by 
\beq 
\langle {\cal Q}_{i}^{(6)} 
\rangle^{\overline{\rm (MS)}}_{\mu_{\overline{\rm MS}}} 
= \left[ \delta_{i j} -
\alpha_s \gamma_{ij} \, 
\ln \left(\frac{\mu^2}{\mu_{\overline {\rm MS}}^{2}}\right) \, +
\alpha_s (f_{ij}^{\overline{\rm MS}} - f_{ij}^{c.o.}) \right] 
\, \langle {\cal Q}_{j}^{(6)} \rangle^{\rm (c.o.)}_\mu 
- \alpha_s {\tilde{\gamma}_{ij} \over \mu^2} \langle 
{\cal O}_{j}^{(8)} \rangle_\mu \ .
\label{connection}
\eeq 

The right hand side of Eq.~(\ref{connection}) is constructed in such a
way to be finite in the limit in which $\mu \rightarrow \infty$. 
The $f_{ij}$ matrices contain finite parts 
which depend upon the particular scheme adopted both in dimensional
regularization and in the cutoff regularization.  In particular,
$f_{ij}^{\overline{\rm MS}}$ depends on the scheme definition 
adopted for $\gamma_5$ away from dimension four.  The coefficients
$\tilde{\gamma}_{ij}$ govern the `leakage' of dimension-eight
operators into the dimension-six sector. They depend on the 
particular scheme adopted for separating scales and, as shown 
before, they can be obtained by a one-loop calculation in the cutoff 
scheme.

\end{document}